\documentstyle[12pt,a4,latexsym,epsf]{article}
\begin{document}

\begin{center}
{\Large Structure of $f_0(980)$ from a Coupled Channel Analysis of 
        $S$-wave $\pi\pi$ Scattering}
\\[5mm]
{\sc M.P.~Locher, V.E.~Markushin and H.Q.~Zheng}
\\[5mm]
{\it Paul Scherrer Institute, 5232 Villigen PSI, Switzerland}
\\[5mm]
August 27, 1997
\\[5mm]
\end{center}

\begin{abstract}
A coupled channel model is used to study the nature of the $f_0(980)$ resonance.
It is shown that the existence of two poles close to $K\bar{K}$ threshold, as
found in many data fits and confirmed here, can be reconciled with
the $K\bar{K}$ molecular origin of the $f_0(980)$. 
Due to a strong coupling between channels the number of the $S$-matrix poles
close to the physical region in the physical case exceeds the number of bare
states introduced in the model.  
\end{abstract}

\section{Introduction}
\label{INT}

The channel with the vacuum quantum numbers $J^{PC}I^G=0^{++}0^+$ has
several resonances \cite{PDT96} the structure of which has been under
discussion for a long time. Recently this problem attracted new attention
after a candidate for the scalar glueball was found at LEAR \cite{CBCg,ACg}.
A characteristic feature of the $J^{PC}I^G=0^{++}0^+$ channel is that the
mixing between different states is strong, so that a connection between the
observed resonances and bare states, like quarkonia or glueballs, is 
nontrivial \cite{To95,An97}. 

  In this paper we shall discuss a part of this problem concerned mainly with
the $f_0(980)$ resonance. This resonance
(the present average values \cite{PDT96} for the mass
and width are: $m_{f_0}=974.1\pm 2.5\;\mbox{\rm MeV}$,
$\Gamma_{f_0}=47\pm 9\;\mbox{\rm MeV}$)
has a long history of experimental and theoretical studies.
The small width and proximity to the $K\bar{K}$ threshold are among its
interesting features, and different explanations have been suggested.
  We list the most prominent ones. A cryptoexotic
$qq\bar{q}\bar{q}$ state was considered in the quark bag model \cite{Ja77}.
A $K\bar{K}$ molecular state was found in the potential quark model
\cite{WI82,GI85,WI90}.
A complicated interplay between $S$-matrix poles and the $K\bar{K}$
threshold structure was shown in \cite{CL86} to play an important role
and obscure conventional simple correspondence between the poles and
resonances.  The phenomenological analysis in the K-matrix framework
favoured for a long time a conventional Breight-Wigner resonance
interpretation (contrary to a $K\bar{K}$ state)
\cite{AMP87,MP93,AS96,AAS97}.  
A quasi-bound $K\bar{K}$ state was found in the coupled channel models
\cite{CDL89,KLM94,KLL97} and in the meson exchange interaction model
\cite{Lo90,JPHS95}.
An exotic vacuum scalar state was considered in \cite{Gr91}. A mixture of
a $q\bar{q}$ state and a scalaron weakly coupled to the $K\bar{K}$ channel was
discussed in \cite{An95}, and a state strongly coupled to the $s\bar{s}$ and
$K\bar{K}$  channels near the $K\bar{K}$ threshold was found in the unitarized
quark model \cite{To95,To96}. 
A state dominated by the $K\bar{K}$ channel was found in a coupled-channel
model derived from the lowest order chiral Lagrangian \cite{OO97}. 
For a more complete list of references dealing with the nature of the $f_0(980)$
see \cite{To95,WI90,MP93,Pa95}.
   
In this paper we shall show that some of the apparently contradicting
interpretations of the $f_0(980)$ resonance represent in fact only a part of
the multifaceted picture of this state resulting from strong coupling between
different channels.  In particular, we revise the conventional claim 
about the connection between the number of poles near the $K\bar{K}$
threshold and the nature of the state ($K\bar{K}$ molecular state vs.
coupled channel resonance, see the discussion in \cite{Mor92} and
references therein).

  Our approach is based on a coupled channel model for the $\pi\pi$ and
$K\bar{K}$ systems. Despite of its simplicity it satisfies the minimal
requirements which allow an adequate phenomenological description of the
$\pi\pi$ scattering amplitudes in the region of the $K\bar{K}$ threshold.
  Since the solutions are available in analytic form, the trajectories of the
S-matrix singularities for coupling constants varied in different ways can
be followed explicitly shedding some light on the physical nature of the
narrow $f_0(980)$ resonance.
  A transparent comparison can be made to similar and alternative models in
the literature.
  The model is introduced in Sec.~\ref{CCM}, and its parameters 
are determined from a fit of the $\pi\pi$ scattering phases. 
  The analytic structure of the scattering amplitudes is presented as a
function of the parameters in Sec.~\ref{POLES}.
  In Sec.~\ref{Disc} the physical properties of the solutions are discussed
and a comparison is made to other approaches in the literature, in
particular the validity of the models relating the $f_0(980)$ resonance to a   
a weakly bound $K\bar{K}$ state is elucidated.
The details of the formalism are collected in the Appendix.

\section{The $\pi\pi-K\bar{K}$ Coupled Channel Model}
\label{CCM}

In order to describe the interaction in the $\pi\pi-K\bar{K}$ system,  we
introduce a simple model with three channels (here and below
we consider the partial wave $J=0^{++}$ $I^G=0^+$). 
Channels 1 and 2 correspond to the $\pi\pi$ and $K\bar{K}$ systems.
Channel 3 consists of a bound state in the $|q\bar{q}\rangle$ channel,
and the rest of the dynamics in this channel is ignored.

The $T$-matrix, as a function of the invariant mass squared $s$, is
defined by the Lippmann-Schwinger equation 
\begin{eqnarray}
    T(s) & = & V + V G^0(s) T
\label{T}
\end{eqnarray}
where $G^0(s)$ is the free Green function 
\begin{eqnarray}
   G^0(s) & = &
   \left(
     \matrix{ G^0_{1}(s) & 0 & 0 \cr
        0 & G^0_{2}(s) & 0 \cr
        0 & 0 & G^0_{3}(s) \cr } 
   \right)
\label{G0}
\end{eqnarray}
The free Green functions for the $\pi\pi$, $K\bar{K}$, and $q\bar{q}$ 
channels have the form
\begin{eqnarray}
  G^0_{1}(s) & = & \frac{2}{\pi}\; 
  \int_{0}^{\infty} \frac{|k_1 \rangle\langle k_1|}{s/4-(m_\pi^2+k_1^2)}
                    k_1^2 dk_1 \\
  G^0_{2}(s) & = & \frac{2}{\pi}\; 
  \int_{0}^{\infty} \frac{|k_2 \rangle\langle k_2| }{s/4-(m_K^2+k_2^2)}
                    k_2^2 dk_2 \\
  G^0_{3}(s) & = &
  \frac{|q\bar{q} \rangle\langle q\bar{q}| }{s-M_r^2}
\label{GGG}
\end{eqnarray}
Here $|k_1\rangle$ and $|k_2\rangle$ denote the free $\pi\pi$ and $K\bar{K}$
states with relative momenta $k_1$ and $k_2$, respectively.
The state $|q\bar{q}\rangle$ in channel 3 has a bare mass $M_r$.

To describe the interaction we use the following potential matrix: 
\begin{eqnarray}
   V & = & 
   \left(
     \matrix{ V_{\pi\pi} & V_{\pi\pi-K\bar{K}}  & V_{\pi\pi-q\bar{q}} \cr
       V_{K\bar{K}-\pi\pi} & V_{K\bar{K}} & V_{K\bar{K}-q\bar{q}} \cr
       V_{q\bar{q}-\pi\pi} & V_{q\bar{q}-K\bar{K}} & 0 \cr } 
   \right)
\label{V}
\end{eqnarray}

We assume that the diagonal interaction $V_{K\bar{K}}$ produces a bound
state in the $K\bar{K}$ channel in the absence of any coupling to the other
channels in our model, thus simulating a `molecular origin' of the $f_0(980)$
resonance\footnote{The effective interaction in the $K\bar{K}$ channel was
demonstrated to be attractive in various quark models 
\cite{To95,WI82,GI85}.}.
A strong coupling of this state to the $\pi\pi$ channel is induced by the
interaction  $V_{\pi\pi-K\bar{K}}=V_{K\bar{K}-\pi\pi}^+$.
The $\pi\pi$ channel is assumed to have a strong coupling to the
$q\bar{q}$ resonance as well. The $q\bar{q}$ state is also directly
coupled to the $K\bar{K}$ channel by the interaction
$V_{K\bar{K}-q\bar{q}}=V_{q\bar{q}-K\bar{K}}^{+}$.  
The diagonal potential $V_{\pi\pi}$ is used to provide a correct description
of the $\pi\pi$ scattering at low energies (see below).

The interaction potentials are taken in separable form:%
\footnote{For our purpose the first rank is sufficient to generate
          the singularities needed.} 
\begin{eqnarray}
   V & = &
   \left( \matrix{
     g_{11}|1 \rangle\langle 1| & g_{12} |1 \rangle\langle 2| &
                                  g_{13} |1 \rangle\langle q\bar{q}| \cr
     g_{12}|2 \rangle\langle 1| & g_{22} |2 \rangle\langle 2| &
                                  g_{23} |2 \rangle\langle q\bar{q}| \cr
     g_{13}|q\bar{q} \rangle\langle 1| &
     g_{23}|q\bar{q} \rangle\langle 2|  & 0 \cr } 
   \right)
\label{Vsep}
\end{eqnarray}
We shall use the following form factors in channel 1 and 2:  
\begin{eqnarray}
   \langle k|1 \rangle & = & \xi_1(k) = \frac{\mu^{3/2}}{k^2+\mu_1^2} \\
   \langle k|2 \rangle & = & \xi_2(k) = \frac{\mu^{3/2}}{k^2+\mu_2^2} 
\label{FF}
\end{eqnarray}
where the parameters $\mu_1$ and $\mu_2$ describe the interaction range
in the $\pi\pi$ and $K\bar{K}$ channels.
 
With this choice of the form factors the matrix elements of the
Green functions are
\begin{eqnarray}
  \langle n | G_n^0(s) | n \rangle & = &
  \frac{\mu_n^2}{2(k_n(s) + i \mu_n)^2} \ \ , \ \ n=1,2 
\label{GME}
\end{eqnarray}
where $k_n(s)$ is the relative momentum in the channel $n$: 
\begin{eqnarray}
   k_1(s) & = & \sqrt{s/4 - m_{\pi}^2} \label{k1s} \\
   k_2(s) & = & \sqrt{s/4 - m_K^2}     \label{k2s} \ \ .
\end{eqnarray}

For our model the analytical solution for the $T$-matrix can be easily
obtained.  The $\pi\pi$ scattering amplitude $f_{\pi\pi}(s)$ has the form:
\begin{eqnarray}
   f_{\pi\pi}(s) & = & 
   - \langle k_1 | T(s) | k_1 \rangle = 
   - \;\frac{\lambda(s) \xi(k_1)^2 }{1 -
            \lambda(s) \langle \xi | G_1^0(s) | \xi \rangle }
\label{f11}
\end{eqnarray}
where
\begin{eqnarray}
   \lambda(s) & = & g_{11} +  g_{13}^2 G_3(s) +  
   \frac{(g_{12}+g_{13}g_{23}G_3(s))^2 \langle 2|G_2^0(k_2)|2\rangle}{1 -
                   (g_{22}+g_{23}^2G_3(s))\langle 2|G_2^0(k_2)|2\rangle}
   \label{ls} \\
       G_3(s) & = & \frac{1}{s - M_r^2} 
\label{T11}
\end{eqnarray}
The first term on the r.h.s. of (\ref{ls}) results from the diagonal
interaction in the $\pi\pi$ channel, the second and third terms correspond to
the effective interactions induced in the $\pi\pi$ channel by the couplings to
the $q\bar{q}$ and $K\bar{K}$ channels.  
We use the coupling constant $g_{11}$ to satisfy the Adler condition
\cite{Ad}, that leads to the vanishing partial wave amplitude 
at $s=m_{\pi}^2/2$, by imposing the following constraint 
\begin{eqnarray}
   \lambda(m_{\pi}^2/2) & = & 0  \ \ . 
\label{AZ}
\end{eqnarray}

The connection between the partial wave $S$-matrix and the scattering
amplitude $f_{\pi\pi}$ is given by
\begin{eqnarray}
     S_{J=0}^{I=0}(s) = \eta_0^0(s) e^{2i\delta_0^0(s)}
          = 1 + 2 i k_1 f_{\pi\pi}(s)
\label{SM}
\end{eqnarray} 
where $\delta_0^0(s)$ is the scattering phase and $\eta_0^0(s)$ is
the elasticity parameter.

\begin{figure}[hbt]
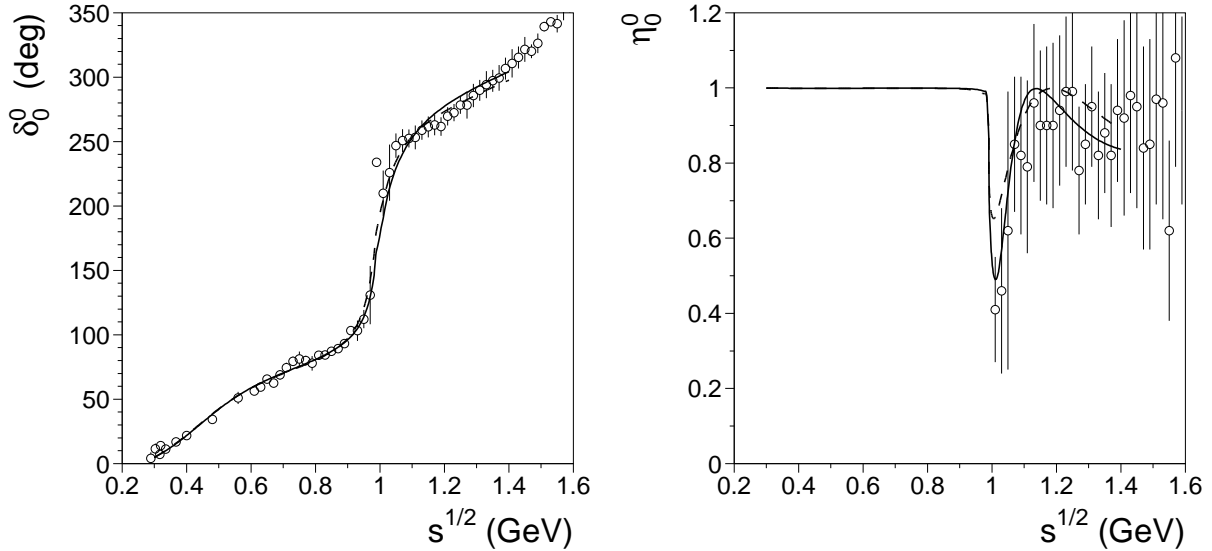

\begin{center}
\mbox{ 
\mbox{\epsfysize=80mm \epsffile{DELTA00.EPSF}} 
\mbox{\epsfysize=80mm \epsffile{ETA00.EPSF}} 
}
\end{center}
\caption{\label{Fit} The scattering phase $\delta_0^0$ and the elasticity
parameter $\eta_0^0$ for the $S$-wave $\pi\pi$ scattering vs. $\sqrt{s}$.
The curves are our model 
(the dashed line --- fit 1, the solid line --- fit 2), 
the experimental points are from
\protect\cite{Gr74,Oc74,Ro77}.}
\end{figure}

\begin{table}[hbt]
\caption{\label{TFit} The model parameters obtained from the fit.
The $g_{11}$ value is calculated using Eq.(\protect\ref{AZ}),
$m_{\pi}=0.1396\;\mbox{\rm GeV}$, $m_K=0.4937\;\mbox{\rm GeV}$.}
\begin{center} \begin{tabular}{|c|ccccccc|c|} \hline
fit & $g_{22}$   & $g_{12}$  & $g_{13}$  & $g_{23}$ 
    & $\mu_1$ & $\mu_2$ & $M_r$  & $g_{11}$  \\
    &            &           & GeV       & GeV  
    & GeV     & GeV     & GeV    &           \\ \hline
 $1^a$   
    & -5.5332  & 3.9456   & 0.69642  &  0
    & 0.37909  & 0.37909    
    & 1.0916   & 4.7499   \\ \hline 
 $2^b$  
    & -3.2421  & 1.9814   & 0.69358  &  0.06828 
    & 0.3730   & 0.89975    
    & 1.0925   & 4.6416    \\ \hline 
\end{tabular} \end{center}
$^{(a)}$ The fit of the $\delta_0^0$ using $g_{23}=0$, $\mu_1=\mu_2$. \\
$^{(b)}$ Including the fit of the $\eta_0^0$ and
         the $K\bar{K}$ scattering data, $\mu_1$ is fixed. 
\end{table}

\begin{figure}[hbt]
\begin{center}
\mbox{\epsfysize=80mm \epsffile{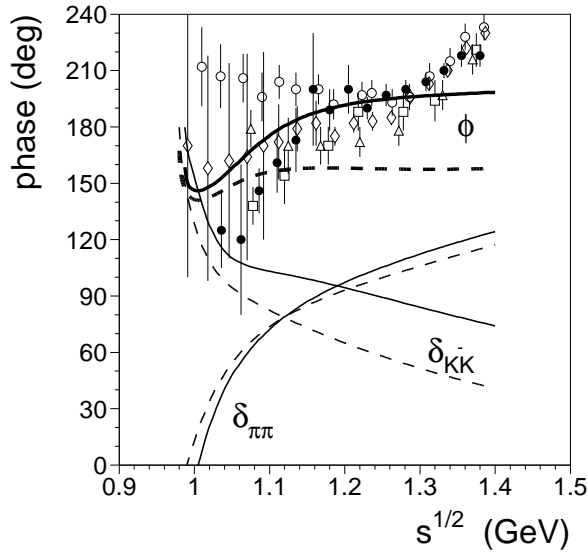}} 
\end{center} 
\caption{\label{DKKbar} The phase shift 
$\phi=(\delta_{\pi\pi}+\delta_{K\bar{K}})$ 
of the $I=0$ $S$-wave scattering $\pi\pi\to K\bar{K}$.    
The results of our model are shown by the thick lines
(fit 1 --- dashed, fit 2 --- solid).   
The data are from
\protect\cite{MaOz} ({\large$\diamond$}), 
\protect\cite{PoZZ} ($\Box$), 
\protect\cite{Co80} ({\large$\circ$}), 
\protect\cite{CoZZ} ($\triangle$),
\protect\cite{Et82} ({\large$\bullet$}). 
The phase shifts for the elastic $S$-wave $I-0$ $\pi\pi$ and $K\bar{K}$
scattering (modulo $180^\circ$) are shown by the thin lines 
(fit 1 ---  dashed, fit 2 --- solid).}
\end{figure}

Considering the coupling constants $g_{22}$, $g_{12}$, $g_{13}$, $g_{13}$, the
interaction ranges $\mu_1$, $\mu_2$ and the position of the bare
$q\bar{q}$ resonance $M_r$ as free parameters, we fitted the $\pi\pi$ scattering
amplitude\footnote{For a recent discussion of the $\pi\pi$ scattering amplitude
see \cite{KLL97,KLR96}.} from \cite{Gr74,Oc74,Ro77} in the energy range
$2m_{\pi} < \sqrt{s} < 1.4\;$GeV.
In fitting the $\pi\pi$ scattering data we found a significant correlation
between the model parameters, in particular, between the coupling constants
and the interaction ranges, so that it was possible to impose some extra 
constraints preserving a good quality of the fit. 
  The fit 1 (see Fig.\ref{Fit} and Table~\ref{TFit}) was done assuming 
equal interaction ranges ($\mu_1=\mu_2$) and switching off the direct
coupling between the $q\bar{q}$ and $K\bar{K}$ channels ($g_{23}=0$). 
  A good fit of the $\pi\pi$ scattering data does not automatically lead
to a good fit of the $K\bar{K}$ scattering (see Fig.\ref{DKKbar}).   
However fitting the $\pi\pi$ and $K\bar{K}$ scattering data together we get
a fair description of the $K\bar{K}$ phase shift as well
(see the fit 2 in Figs.\ref{Fit},\ref{DKKbar} and Table~\ref{TFit}).
  The calculated $S$-wave $K\bar{K}$ scattering phase shown in
Fig.\ref{DKKbar} has an energy behaviour typical for the presence of a
weakly bound state.  
  The two fits shown represent typical results obtained with our model. 
Because of a significant uncertainty in the data near the $K\bar{K}$
threshold%
\footnote{For a discussion of constraining the $\pi\pi-K\bar{K}$ scattering
amplitude by extending the set of experimental data,
see \cite{AMP87,MP93,AS96,JPHS95}.}  
and the importance of higher resonances which are not considered here
we shall use both sets of the model parameters in the further analysis,
the fit 2, however, appears to be preferable.

\section{The Poles of the $S$-Matrix}
\label{POLES}

\subsection{The Analytical Structure of the $S$-Matrix}
\label{ASSM}

\begin{figure}[hbt]
\begin{center}
\mbox{\epsfysize=4cm \epsffile{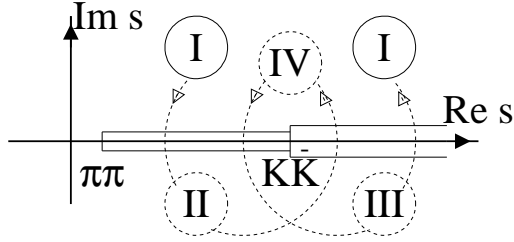}} \vspace*{-15mm}
\end{center}
\caption{\label{RS} The Riemann surface of the scattering amplitude.}
\end{figure}

The Riemann surface of the scattering amplitude for the two channel problem
has four sheets due to the kinematical cuts starting at the $\pi\pi$ and
$K\bar{K}$ thresholds according to Eqs.(\ref{k1s},\ref{k2s}), as shown in
Fig.\ref{RS}.
The sheets of the complex $s$-plane are distinguished by the signs of the
imaginary parts of the channel momenta $k_1$ and $k_2$, with the standard
notation given in Table~\ref{TRS}.
The physical scattering region corresponds to the upper side of the cut
going along the real $s$ axis on the sheet I. If there exist true bound
states in both channels, they occur on the sheet I on the real $s$-axis
below all the thresholds.    
The sheet II accommodates the states which appear as resonances in the
channel 1 ($\pi\pi$) and (quasi) bound states in channel 2 ($K\bar{K}$). The
sheet III corresponds to resonances in both channels.    

\begin{table}[hbt] 
\caption{\label{TRS}
The sheets of the Riemann surface of the scattering amplitude
in the $s$-plane.}
\begin{center} \begin{tabular}{c|cc} \hline
 Sheet & \mbox{Im $k_1$} & \mbox{Im $k_2$} \\ \hline
    I  &   +   &   +  \\
   II  &  $-$  &   +  \\
  III  &  $-$  &  $-$  \\
   IV  &   +   &  $-$ \\ \hline 
\end{tabular} \end{center}
\end{table} 

Using our model we found the position of the poles of the $S$-matrix in the
complex $s$-plane, the result being shown in Table~\ref{TPole}.
There are two poles, $s_A^{II}$ and $s_D^{III}$, very close to the
$K\bar{K}$ threshold. Another pair of poles, 
$s_B^{II}$ and $s_E^{III}$, corresponds to a broad resonance above
the $K\bar{K}$ threshold, and the pole $s_C^{II}$ corresponds to a broad
structure associated with the $\pi\pi$ threshold. The physics of these poles
will be discussed in detail in Sec.\ref{POLETR}.
The position of the resonance poles depends on the coupling
constants, and this distinguishes them from the fixed poles originating
from the singularities of the form factors (\ref{FF}). The latter are located
at $k_1=\pm i\mu_1$ and $k_2=\pm i\mu_2$, their proximity to the physical
region being determined by the range of the interaction. In our model
these fixed poles approximate the potential singularities which correspond 
to the left hand cut in a more general case. 

\begin{table}[hbt]
\caption{\label{TPole}
The resonance poles of the $S$-matrix in the $s$-plane (GeV$^2$).}
\begin{center} \begin{tabular}{c|c|c|c} \hline
 Pole & Sheet &  fit 1            &  fit 2            \\
\hline
   A  & II    &  $0.979 - i0.065$ & $1.015 - i0.054$  \\ 
   B  & II    &  $1.653 - i1.163$ & $1.521 - i1.092$  \\ 
   C  & II    &  $0.135 - i0.180$ & $0.131 - i0.178$  \\ 
   D  & III   &  $1.007 - i0.158$ & $1.072 - i0.193$  \\ 
   E  & III   &  $1.562 - i1.773$ & $1.207 - i1.696$  \\ 
\hline
\end{tabular} \end{center}
\end{table} 

\subsection{Trajectories of the Resonance Poles}
\label{POLETR}

To understand the origin and the nature of the resonance poles found in our
model we investigate how these poles move in the complex $s$-plane when the
model parameters are varied between the physical case and the limit of  
vanishing couplings between the $\pi\pi$, the $K\bar{K}$, and the $q\bar{q}$
channels:  $g_{12}=g_{13}=0$. In this
case the diagonal interaction  in the $K\bar{K}$ channel with the {\it
physical} strength of the coupling $g_{22}$  produces a bound state close to
the $K\bar{K}$ threshold with mass $m_{K\bar{K}}=0.85\;$GeV
(for the fit 2 the coupling with the $q\bar{q}$ channel contributes
$8\;$MeV to the binding energy).  
The $q\bar{q}$ state in the absence of coupling to the open channels is
characterized by the mass $M_r$ and zero width.    

\begin{figure}
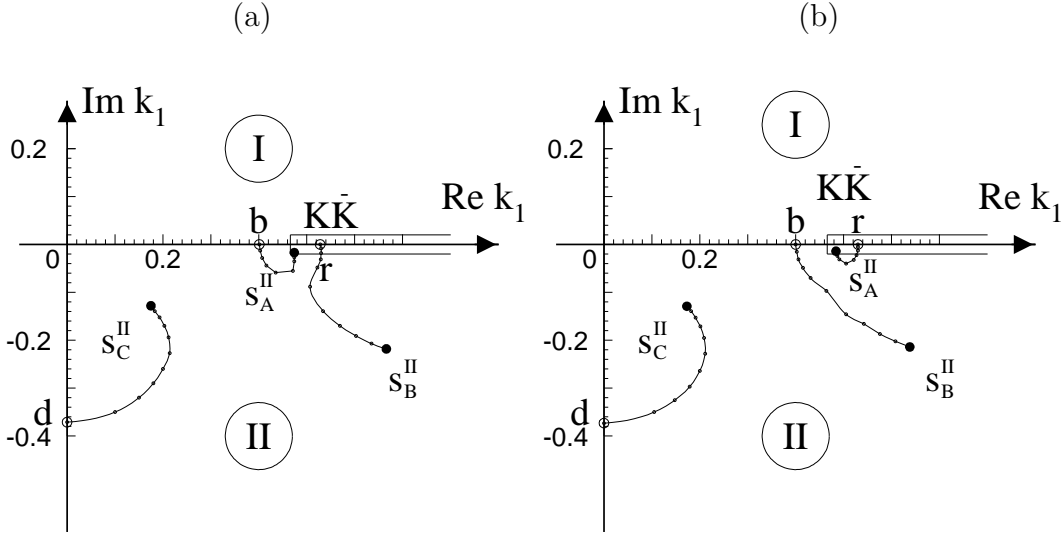

\begin{center}
\mbox{(a)}\hspace*{70mm}\mbox{(b)} \\
\mbox{ 
\mbox{\epsfysize=70mm\epsffile{VM2PANIC05A.EPSF}}
\mbox{\epsfysize=70mm\epsffile{CCM242K1.EPSF}}
} 
\end{center}
\vspace*{-5mm} 
\caption{\label{FPk1}
The trajectories of the poles in the complex $k_1$ plane  for the
$K\bar{K}-\pi\pi$ and $q\bar{q}-\pi\pi$ couplings increasing from $x=0$ 
($\circ$) to the physical values $x=1$ ($\bullet$).
The labels indicate the original positions of the bound state $(b)$,
the $q\bar{q}$ resonance $(r)$, and the dynamical pole $(d)$.
The dots on the trajectories mark the increase of $x$ in steps of $0.1$.
Fit 1 is shown in Fig.(a), fit 2 in Fig.(b).}
\end{figure}

\begin{figure}
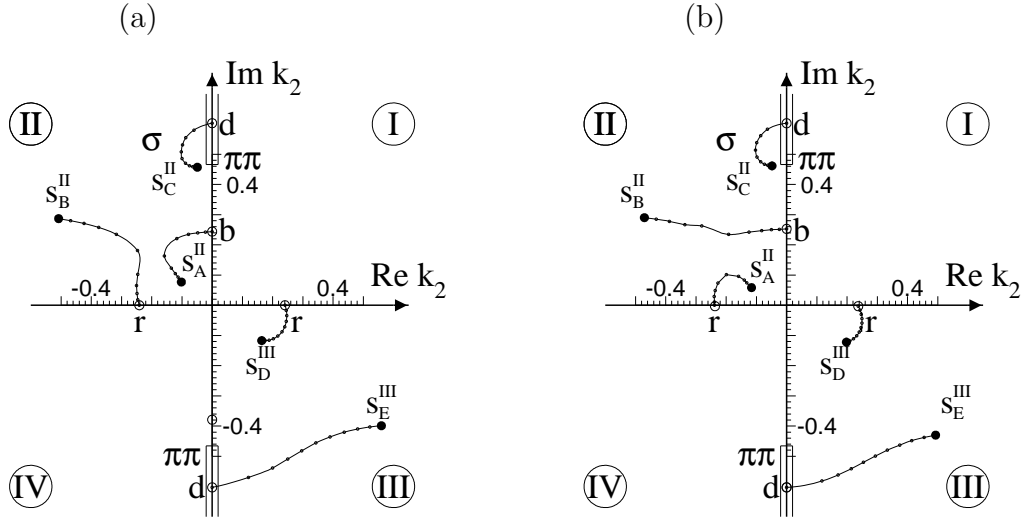

\begin{center}
\mbox{(a)}\hspace*{70mm}\mbox{(b)} \\
\mbox{ 
\mbox{\epsfysize=75mm\epsffile{VM2PANIC06A.EPSF}}
\mbox{\epsfysize=75mm\epsffile{CCM242K2.EPSF}}
} 
\end{center}
\vspace*{-5mm} 
\caption{\label{FPk2}
The trajectories of the poles in the complex $k_2$ plane  for the
$K\bar{K}-\pi\pi$ and $q\bar{q}-\pi\pi$ couplings increasing from 0 ($\circ$) to
the physical values ($\bullet$). 
Fit 1 is shown in Fig.(a), fit 2 in Fig.(b). 
The labels are as in Fig.\protect{\ref{FPk1}}.}
\end{figure}

  The trajectories of the $S$-matrix poles were calculated for the coupling
constants $g_{12}$ and $g_{13}$ varied between zero and the physical
values given in Table~\ref{TFit}:
\begin{eqnarray}
    g_{12} & \to & x^{1/2} g_{12} \ , \nonumber \\     
    g_{13} & \to & x^{1/2} g_{13} \ , \ \  0 \leq x \leq 1
\label{xgg}     
\end{eqnarray}
Figures \ref{FPk1} and \ref{FPk2} display the trajectories plotted in
the complex planes of the channel momenta $k_1$ and $k_2$.
The four sheets of the complex $s$-plane correspond to two sheets of the
complex momenta planes. Figure~\ref{FPk1} shows the $k_1$ plane that
contains the sheets I and II of the $s$-plane (the second sheet of the
$k_1$ plane, which is not shown here, is reached across the cut going from
the $K\bar{K}$ threshold to infinity).
All four sheets of the $s$-plain can be displayed on a single sheet of the
$k_2$ plane with the kinematical cuts going from the $\pi\pi$ thresholds to
infinity, as shown in Figure~\ref{FPk2}. The second sheet in the $k_2$ plane
is a mirror copy (with respect to the imaginary $k_2$ axis) of the plotted
one due to the symmetry properties of the $S$-matrix \cite{SSym}.

 There are three poles on the sheet II.
The pole trajectories for the fits 1 and 2 shown in Fig.\ref{FPk1} 
demonstrate two different cases of interplay between the original
$K\bar{K}$ bound state and the $q\bar{q}$ resonance. 
For the fit 1, the pole $s_A^{II}$  originates
from the bound $K\bar{K}$ state and develops an imaginary part due to the
coupling to the $\pi\pi$ channel, see the trajectory $(b-S_A^{II})$ in
Figs.\ref{FPk1}a,\ref{FPk2}a.   This pole remains, however, close to the
$K\bar{K}$ threshold and does not go far away from the real $k_1$ axis. This
occurs because of its interplay with the pole $s_B^{II}$ which belongs to 
the trajectory $(r-s_B^{II})$ originating from the $q\bar{q}$ state.
At small nonzero couplings (\ref{xgg}) these two poles
develop comparable imaginary parts. With increasing coupling the two
poles first get closer but then they start to move away from each other. As
a result the pole $s_B^{II}$ gets a large imaginary part while the pole
$s_A^{II}$ turns back to the real axis.
  This kind of pole motion is one of the generic possibilities occurring in
the problem of two states coupled via a continuum as discussed in
the Appendix.

  The fit 2 corresponds to the second generic possibility when the pole
originating from the $q\bar{q}$ resonance is attracted to the $K\bar{K}$
threshold producing the narrow state (see the trajectory $(r-s_A^{II})$ 
on Figs.\ref{FPk1}b,\ref{FPk2}b) while the pole originating from the
$K\bar{K}$ bound state develops a large width due to the strong coupling to
the $\pi\pi$ channel.
  In both cases the strong attractive interaction in the $K\bar{K}$
channel is essential for the appearance of the pole on the sheet II
close to the $K\bar{K}$ threshold. 

  In both cases the pole $s_C^{II}$ is generated {\it dynamically} by the
effective attractive interaction in the $\pi\pi$ channel resulting from 
the coupling to the closed channels. The dynamical nature of this pole is
seen from the fact that its trajectory $(d-S_C^{II})$ begins at the
singularity of the form factor $\xi(k_1)$ at $k_1=-i\mu_1$.
  This very broad resonance is responsible for the strongly attractive $\pi\pi$
scattering phase between the $\pi\pi$ and $K\bar{K}$ thresholds and can be
associated with the $\sigma$ meson found in other models
\cite{To95,KLM94,KLL97,JPHS95,To96,OO97,BL71,ZB94,AS94}.
The mass and the width of the $\sigma$ meson is obtained 
using the pole position, $s_C^{II}=(M_{\sigma}-i\Gamma_{\sigma}/2)^2$,  
the fits 1 and 2 give similar results:
$M_{\sigma} \approx 0.42\;$GeV and $\Gamma_{\sigma} \approx 0.42\;$GeV.

  The sheet III contains two pole trajectories showing a similar behaviour
in both cases. 
The $q\bar{q}$ state, being a genuine coupled channel resonance, corresponds to
two poles (in the zero coupling limit they are symmetric with respect to the
imaginary $k_2$ axis), see Fig.\ref{FPk2}. The pole on the sheet II 
was considered above. The pole on the sheet III moves along
the trajectory $(r-s_D^{III})$ closer to the $K\bar{K}$ threshold with
increasing coupling to the open channels. For the fit 1, at the physical
values of the coupling constants the pole $s_D^{III}$ positions itself 
{\it seemingly} as a partner of the pole $s_A^{II}$ with {\it small} imaginary
part, while, by its origin, the pole $s_D^{III}$ is a counterpart of
the pole $s_B^{II}$, as is clearly seen in the weak coupling regime.
  For the fit 2, the poles moving along the trajectories originating from the
initial $q\bar{q}$ poles, $(r-s_A^{II})$ and $(r-s_D^{III})$, look like 
counterparts for all values of the channel coupling constant between zero
and the physical value. The attractive $K\bar{K}$ interaction
is essential for keeping these poles close to the $K\bar{K}$ threshold,
which again speaks for the $K\bar{K}$ nature of the $f_0(980)$.   
  This result illustrates that on the basis of the physical location of the
poles alone the nature of the resonances in strong coupling limit cannot be
determined reliably. The study of the {\it trajectories} in a model can be
helpful for uncovering the underlying dynamics. 

The pole $s_E^{III}$ on the sheet III has a very large imaginary part, thus
it looks like a counterpart of the pole $s_B^{II}$ on the sheet II, as
expected for the broad $q\bar{q}$ state strongly coupled to the open channels. 
However, this naive interpretation  cannot be correct because the pole
originating from the $q\bar{q}$ state on the sheet III was found to be
attracted to the $K\bar{K}$ threshold and has a small width. Figure \ref{FPk2}
shows that the pole $s_E^{III}$ has a dynamical origin: the corresponding
trajectory emerges from the dynamical singularity of the effective
attractive potential in the $\pi\pi$ channel ($k_1=-i\mu_1$).  
This result holds for both fits. 

The existence of nearby resonance poles of the S-matrix which are 'far away'
in the limit of zero channel coupling is a characteristic
feature of coupled channel models with potential (left hand) singularities.
That is why the coupled channel model can provide a picture of the pole
trajectories different from the standard K-matrix parametrization where no
poles are generated dynamically, such that all poles must be introduced
explicitly.   

\begin{figure}[htb]
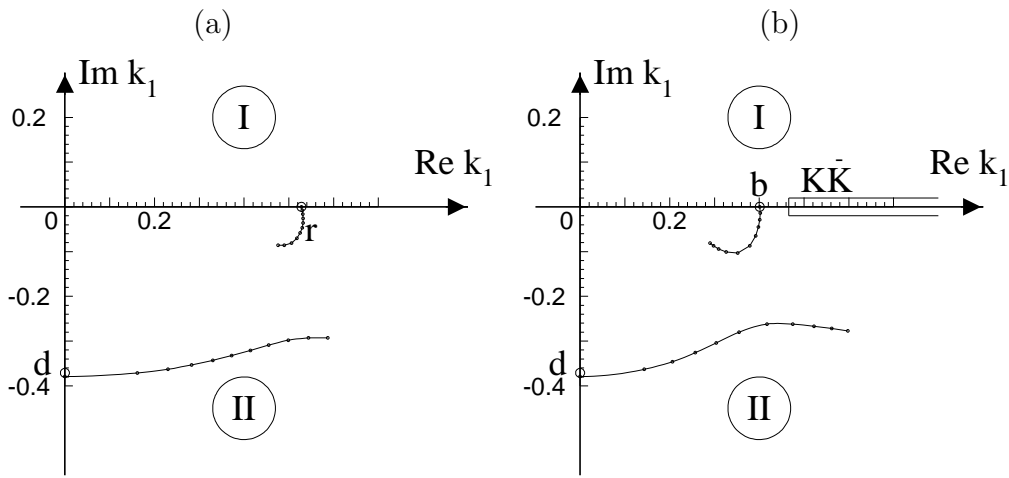

\begin{center}
\mbox{(a)}\hspace*{70mm}\mbox{(b)} \\
\mbox{\epsfysize=60mm \epsffile{K1VAR13.EPSF}}
\mbox{\epsfysize=60mm \epsffile{K1VAR12.EPSF}}
\end{center}
\caption{\label{k1var}
The trajectories of the poles in the complex $k_1$ plane when one of the
channels is switched off:
(a) the $q\bar{q}$ state is coupled only to the $\pi\pi$ channel 
$(g_{12}=g_{22}=0)$; 
(b) the $\pi\pi$ and $K\bar{K}$ channels are coupled to each other without
coupling to the $q\bar{q}$ state $(g_{13}=0)$,
all other parameters correspond to the fit 1.  
The labels are as in Fig.\protect{\ref{FPk1}}. }
\end{figure}

The interaction between the $K\bar{K}$ bound state and the $q\bar{q}$ state is
important for producing the narrow $f_0(980)$ resonance. This can be
illustrated by comparing the pole trajectories in Fig.\ref{FPk1} with the ones
calculated for the cases when either the $K\bar{K}$ or the $q\bar{q}$ channel
is switched off, see Fig.\ref{k1var}.
If the $q\bar{q}$ state is coupled only to the $\pi\pi$ channel, then
at the physical strength of the coupling $g_{13}$ the resonance
originating from the $q\bar{q}$ state has a width of about
$330\;\mbox{\rm MeV}$, $s_r=(0.96-i0.33)\;\mbox{\rm GeV}^2$.
If the $K\bar{K}$ channel is coupled only to the $\pi\pi$ channel, then
the $K\bar{K}$ state at the physical strength of the coupling $g_{12}$
has a width of about $300\;\mbox{\rm MeV}$,
$s_b=(0.39-i0.19)\;\mbox{\rm GeV}^2$. 
In both cases an additional very broad resonance with a width of about
$1\;\mbox{\rm GeV}$ arises from the dynamical singularity as a result of
the strong coupling between the channels.    

\section{Discussion}  \label{Disc}

  In this section we discuss how our results are related to previous
studies in the literature.
  In our model the final configuration representing the data has a {\it pair}
of poles close to the $K\bar{K}$ threshold.
The pole $s_A^{II}$ on sheet II corresponds to a strong interplay between
the original $K\bar{K}$ bound state and the $q\bar{q}$ resonance, the
attractive $K\bar{K}$ interaction being important for its proximity to the
threshold. Both generic possibilities of the interaction between two poles
(see Appendix~A) were found to be consistent with the $\pi\pi$ scattering
data.
When our model is constrained by the $\pi\pi-K\bar{K}$ scattering data,  
the rearrangement situation is favoured (fit 2):
with increasing coupling to the $\pi\pi$
channel the pole originating from $K\bar{K}$ (molecular) bound state nearly
collides with the one originating from the $q\bar{q}$ state and then
it is repelled and develops a large imaginary part while the pole
originating from the $q\bar{q}$ state is attracted to the $K\bar{K}$
threshold. However the alternative situation (fit 1),
when the pole $s_A^{II}$
belongs to the trajectory originating from the $K\bar{K}$ bound state,
can be realized with relatively small variation of the model parameters and
thus is as well a viable solution. 

  Next in importance by distance to the physical region is the pole
$s_D^{III}$ on sheet III, which arises from the $q\bar{q}$ resonance
strongly coupled to the $\pi\pi$ and $K\bar{K}$ channels.
  The remote pole $s_B^{II}$ on the sheet II has a dynamical counterpart
$s_E^{III}$ with large width, the latter is created by the interaction
between the $q\bar{q}$ state and the $\pi\pi$ channel on sheet III.
  
This configuration of the poles is similar to the result found for the 
fits in the K-matrix formalism \cite{AMP87,MP93}.
The existence of {\it two}
poles close to the $K\bar{K}$  threshold  was often interpreted as an evidence
against the $K\bar{K}$ molecular state origin of the $f_0(980)$ resonance.
While in the limit of weak coupling to the $\pi\pi$ channel the $K\bar{K}$
bound state is clearly described by a single pole, the situation can be very
different in the case of strong coupling as our model explicitly
demonstrates.       

It was found in \cite{MP93} that the amplitude in the $K\bar{K}$ molecular
model of the $f_0(980)$ \cite{WI90} has only one pole near the $K\bar{K}$
threshold. This, however, does not mean that any other model with the
$K\bar{K}$ weakly bound state has the same feature, and our model is just a
counterexample. As demonstrated in \cite{MP93}, the fits, which allow two
poles near the $K\bar{K}$ threshold, describe the data much better than an
one-pole fit. Because in our model the number of the resonance poles is not
limited by the number of the bare states, the second pole needed for the
best fit to the data is generated dynamically and the overall picture is
similar to the favoured two-pole solution in \cite{MP93}.  

The exact position of the poles corresponding to the $f_0(980)$ resonance
is known to be sensitive to the selection of data used in the fit.
Our results concerning the mass and the width of the $f_0(980)$ are compared 
with some recent results in Table~\ref{Tfo}
(for detailed description of input data we refer to the original papers).   

\begin{table}[hbt]
\caption{\label{Tfo}
The of the $S$-matrix in the $\sqrt{s}$-plane (GeV)
corresponding to the $f_0(980)$ resonance.}
\begin{center} \begin{tabular}{c|cc} \hline
 Ref. & $M_{f_0}-i\Gamma_{f_0}/2\ $ (sheet II) & $M_{f_0}-i\Gamma_{f_0}/2\ $
                                                 (sheet III) \\ \hline
  fit 1       & $0.990 - i0.033$        & $1.007 - i0.079$ \\
  fit 2       & $1.008 - i0.027$        & $1.039 - i0.093$ \\
 \cite{CBCg}  & $0.996 - i0.056$        & $0.953 - i0.055$ \\
 \cite{To95}  & $1.006 - i0.017$        &    \\
 \cite{AMP87}$^a$ 
              & $1.001 - i0.026$        & $0.985 - i0.020$ \\[-2mm] 
              & $0.988 - i0.000$        &    \\ 
 \cite{MP93}$^b$  
              & $0.988 - i0.024$        & $0.978 - i0.028$ \\
 \cite{MP93}$^c$
              & $0.972 - i0.016$        &    \\
 \cite{AS96}  & $1.008 - i0.043$        & $0.957 - i0.041$ \\
 \cite{CDL89} & $0.993 - i0.023$        &    \\
 \cite{KLM94} & $0.973 - i0.029$        &    \\
 \cite{KLL97}$^d$ 
              & $0.989 - i0.031$        &    \\[-2mm]
              & $0.992 - i0.034$        &    \\ 
 \cite{JPHS95}& $1.015 - i0.015$        &    \\ 
 \cite{An95}  & $0.987 - i0.040$        & $0.967 - i0.069$ \\
 \cite{ZB94}  & $0.988 - i0.023$        & $0.797 - i0.185$ \\
 \cite{ABSZ94}& $0.984 - i0.039$        & $0.986 - i0.102$ \\
 \cite{An96}  & $1.015 - i0.043$        &    \\
\hline
\end{tabular}
\\[2mm]
\parbox{100mm}{\small
a) The three-pole fit. \\ 
b) The favoured two-pole fit. \\ 
c) Using the model from \protect\cite{WI90}. \\ 
d) Fits to the ``down-flat'' and ``up-flat'' data. } 
\end{center}
\end{table} 

The dynamical description of the $f_0(980)$ by two poles in our model is similar
to the result of the unitarized quark model \cite{To95} where the strong
coupling between the bare quark-antiquark states and the two meson channels
was found to produce additional poles. Our model is different from
\cite{To95} concerning the analytical structure of the transition
form-factors, see Eq.(\ref{FF}).
In our case the form-factors have poles whose positions are 
related to the range of interaction. In the model \cite{To95} the
form-factors are taken in the oscillator form and have an essential singularity
at infinity. The shape of the form factors does not seem to be crucial for
the quality of fitting the data, but our choice has the advantage of placing 
the leading dynamical singularities at a finite distance from
the thresholds.
 As both models demonstrate, the interplay between the bare states and the
dynamical singularities is important for the correct phenomenological
description of the resonances in the scalar-isoscalar channel. An essential
feature is that the number of resonance poles can exceed the
number of the bare states. 

\begin{table}[hbt]
\caption{The poles of the $S$-matrix in the $\sqrt{s}$-plane (GeV)
         corresponding to the $\sigma$ resonance.}
\label{Tsigma} 
\begin{center} \begin{tabular}{c|c} \hline
 Ref.         & $M_{\sigma}-i\Gamma_{\sigma}/2\ $ (sheet II) \\ \hline
 fit 1        & $0.424 - i0.213$ \\
 fit 2        & $0.420 - i0.212$ \\
 \cite{To95,To96}
              & $0.47  - i0.25 $ \\
 \cite{KLM94} & $0.506 - i0.247$ \\
 \cite{KLL97}$^a$ 
              & $0.518 - i0.261$ \\[-2mm] 
              & $0.562 - i0.233$ \\  
 \cite{JPHS95}& $0.387 - i0.305$ \\
 \cite{OO97}  & $0.470 - i0.179$ \\  
 \cite{BL71}  & $0.460 - i0.338$ \\
 \cite{ZB94}  & $0.370 - i0.356$ \\
 \cite{AS94}  & $0.42  - i0.37 $ \\
\hline
\end{tabular}
\\[2mm]  
\parbox{100mm}{\small a) Fits to the ``down-flat'' and ``up-flat'' data. } 
\end{center}
\end{table} 

The $\sigma$ meson is generated dynamically as a broad coupled channel 
resonance and hence corresponds to a smooth attractive background phase. The
parameters of this resonance are expected to be model dependent since the 
very existence of a broad resonance cannot be proven on the basis of data 
alone.  Our results for the $\sigma$ are compared with the values from the 
literature in Table~\ref{Tsigma}. 

  There is a broad resonance above the $K\bar{K}$ threshold, which is related
to the $q\bar{q}$ state strongly coupled with the $\pi\pi$ and $K\bar{K}$ 
channels, as discussed above. The mass and the width calculated from the
position of the poles $s_B^{II}$ abd $s_E^{III}$ are 
$(M_B-i\Gamma_B/2)=(1.36-i0.43)\;\mbox{\rm GeV}$ and
$(M_E-i\Gamma_E/2)=(1.40-i0.63)\;\mbox{\rm GeV}$ (the fit 1).  
This resonance is similar to that of the K-matrix analysis \cite{MP93} where 
a very broad state at about $1\;\mbox{\rm GeV}$ of width around
$0.7\;\mbox{\rm GeV}$ was found.
In our model this resonance is needed to describe the rise of the $\pi\pi$
scattering phase above the $K\bar{K}$ threshold. 
It can be considered as an approximation for the higher resonances,
in particular, the $f_0(1300)$, and our results for the remote poles
$s_D^{II}$ and $s_E^{III}$ should be considered as an estimate rather
than the parameters of real physical structures  (for the discussion
of this energy region see \cite{ACg,To95,An97,An96,KLL97,KlLo97} and
references therein).  

A qualitative description of this mass range would require accounting for
additional higher resonances, see \cite{An97,ABSZ94,An96}.    
We do not attempt it here, however, we can use our model to demonstrate how 
the $q\bar{q}$ state dissolves into the continuum. This problem is especially
interesting because the channel coupling is strong, such that the poles
corresponding to the very broad resonance and the bare state belong to
different trajectories (compare the $s_E^{III}$ and the $s_D^{III}$ in
Fig.\ref{FPk2} in both fits). Our approach is based on the probability sum
rule for a resonance embedded into a continuum \cite{BHM}. Using the full 
Green function
\begin{eqnarray}
    G(s) & = & G^0(s) (1 + T(s) G^0(s)) 
\end{eqnarray}
and projecting the completeness relation in channel space 
onto the $q\bar{q}$ channel one gets
\begin{eqnarray}
   \int_{4 m_{\pi}^2}^{\infty} w(s) ds
    & = & \langle q\bar{q}|q\bar{q} \rangle = 1 \label{SumR} \ \ , 
\end{eqnarray} 
where
\begin{eqnarray}
   w(s)   & = & \frac{1}{2\pi i}(G_3(s-i\epsilon)-G_3(s+i\epsilon)) \ \ ,
   \label{w} \\
   G_3(s) & = & \langle q\bar{q}|G(s)|q\bar{q} \rangle 
                      = \frac{1}{s - M_r^2 - \Pi(s)}  \ \ , \label{G3} 
\end{eqnarray}
and $\Pi(s)$ is the mass operator of the $q\bar{q}$ state:
\begin{eqnarray}   
   \Pi(s) & = &
   \frac{4
   (g_{13}^2 G_1^0(s) + g_{23}^2 G_2^0(s) + 
   (2g_{12}g_{13}g_{23} - g_{13}^2g_{22} - g_{23}^2g_{11}) G_1^0(s)G_2^0(s))}
   {1 - g_{11} G_1^0(s) - g_{22} G_2^0(s) +
       (g_{11}g_{22}-g_{12}^2) G_1^0(s) G_2^0(s) }
   \ \ . 
\label{MOG3}
\end{eqnarray}

\begin{figure}[htb]
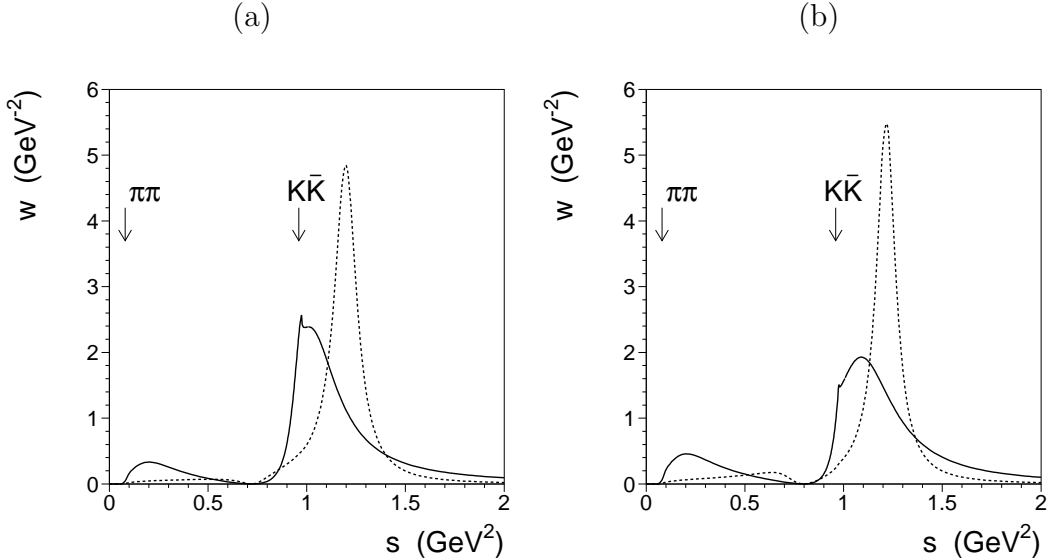

\begin{center}
\mbox{(a)}\hspace*{70mm}\mbox{(b)} \\
\mbox{
\mbox{\epsfysize=70mm \epsffile{SR120.EPSF}}
\mbox{\epsfysize=70mm \epsffile{SR242.EPSF}}
}
\end{center}
\caption{\label{SR}
The probability density for the $q\bar{q}$ state embedded in the continuum.
The solid line corresponds to the physical parameters,
the dashed line is for $x=0.2$ in Eq.(\protect\ref{xgg}):
(a) fit 1, (b) fit 2.}
\end{figure}

According to Eq.(\ref{SumR}) the function $w(s)$ determines the probability
density for the $q\bar{q}$ component in the scattering states within
the interval $(s,s+ds)$, as shown in Fig.\ref{SR}. 
In the case of weak coupling the probability density is well localized near
the position of the bare $q\bar{q}$ state. For the physical case we find a
broad peak centered slightly above the $K\bar{K}$ threshold, so that 
there is a significant overlap of the $w(s)$ distribution with the $f_0(980)$
resonance. This down shift with respect to the position of the original
$q\bar{q}$ state is due to the strong attraction of the $s_D^{III}$ pole 
towards the $K\bar{K}$ threshold. 
  The position and the width of the $w(s)$ distribution indicates that an
essential contribution to the saturation of the sum rule (\ref{SumR}) comes
from the pole $s_D^{III}$, while the narrow structure associated with 
the pole $s_A^{II}$ alone plays a minor role.
Thus out of the two poles related to the narrow $f_0(980)$ state
only the pole on sheet III has a large $q\bar{q}$ component and the pole
on sheet II has mainly $K\bar{K}$ nature.

\section{Conclusion}  \label{Concl}

 We have re-examined the resonance structures in the $J^{PC}I^G=0^{++}0^+$
partial wave of $\pi\pi$ scattering  below 1 GeV, including the $f_0(980)$ and
$\sigma$ resonances on the basis of an exactly solvable coupled channel
model.   The model has the following features: a separable diagonal
$K\bar{K}$ potential producing a weakly bound state, a separable transition
potential $V_{\pi\pi-K\bar{K}}$ (representing $K^*$-exchange, e.g.) which
couples the $\pi\pi$ and $K\bar{K}$ channels, and a broad resonance in the
$q\bar{q}$ channel which represents background and is coupled to
the $\pi\pi$ and $K\bar{K}$ channels.   
  Tuning the model parameters the experimental energy dependence of the
$\pi\pi$ $S$-wave scattering phase $\delta_{0}^{I=0}(s)$ is 
reproduced accurately.
  The interpretation of the singularities is elucidated by tracing the 
trajectories of the $S$-matrix poles as a function of the  strength of the
channel couplings. 
  Two generic cases are represented by fit~1 and fit~2. 

  In our model the $f_0(980)$ resonance corresponds to two $S$-matrix poles
close to the $K\bar{K}$ threshold.
The pole on sheet II ($s_A^{II}$) corresponds to the interplay between the
original $K\bar{K}$ (molecular) bound state and the $q\bar{q}$ state due to
their coupling via the $\pi\pi$ channel. This pole has many features
typical for the $K\bar{K}$ bound state in the absence of the coupling
to the $\pi\pi$ channel. The model parameters consistent with the data
allow two possibilities when the coupling with the $\pi\pi$ channel is
switched on: the pole $s_A^{II}$ originates either directly
from the $K\bar{K}$ bound state (fit 1) or through the rearrangement from
the $q\bar{q}$ state colliding with the $K\bar{K}$ state (fit 2).      
  The second pole on sheet III ($s_D^{III}$) arises from the $q\bar{q}$
resonance coupled to the $\pi\pi$ and $K\bar{K}$ channels. It has a large
$q\bar{q}$ component, but its position close to the $K\bar{K}$ threshold is
due to the attractive interaction in the $K\bar{K}$ channel. 
There is no contradiction between the presence of two poles close to
$K\bar{K}$ threshold and the molecular origin of the $f_0(980)$ resonance.

  The remote poles on sheet II ($s_B^{II}$)  and sheet III ($s_E^{III}$)
correspond to a smooth background phase, the latter originates from the
dynamical singularity describing the coupling of the $q\bar{q}$ resonance to
the $\pi\pi$ channel. 
  The $\sigma$ meson is generated dynamically by the strong coupling 
$q\bar{q} \leftrightarrow \pi\pi$ resulting in an effective attractive
interaction in the $\pi\pi$ channel below the $q\bar{q}$ state ($s_C^{II}$).   
  The appearance of dynamical coupled channel poles is a crucial
feature; in the physical case the number of poles close to the physical region
exceeds the number of bare states in the model.  

  Overall the resulting configuration is similar to the results from the
phenomenological K-matrix analysis: two poles close to the $K\bar{K}$
threshold. Our model shows that the interpretation of the scattering data is
entirely consistent with a $K\bar{K}$ state picture for the $f_0$
resonance, which is thus not in conflict with the (phenomenological)
requirement of having two poles (not one) near the $K\bar{K}$ threshold.

\section*{Acknowledgments}

We thank M.R.~Pennington for comments on the nature of the $f_0(980)$ and
L.~Le\'sniak for a discussion of the $\pi\pi - K\bar{K}$ scattering data.

\section*{Appendix A. Two states coupled via a continuum}

Contrary to direct coupling between two states which is well known to result
in their repulsion, a coupling of two states via a continuum channel
can induce an effective attraction.
This problem has been treated in the literature (see, for example,
\cite{TSCC}), so here we give only a brief summary for the benefit of
the reader. 

\begin{figure}[hbt]
\begin{center}
\mbox{\epsfxsize=40mm \epsffile{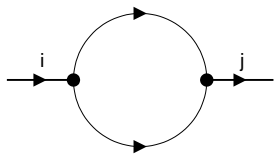}} 
\end{center}
\caption{\label{TSC} The system of two states $(i,j=1,2)$ 
coupled with the continuum channel.}
\end{figure}

\begin{figure}
\begin{center}
\mbox{\epsfysize=60mm \epsffile{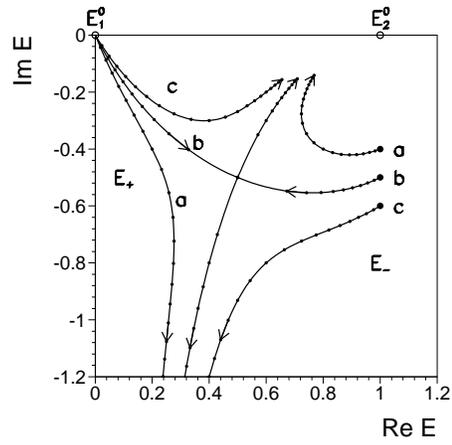}}
\end{center}
\caption{\label{TSP}
The poles of the Green function in the complex energy
plane for the system of two states coupled via a continuum.
The trajectories show the motion of the poles as function of 
$\Gamma_1$ for three different cases:
(a) $\Gamma_2 = 0.8 |E_1^0 - E_2^0|$,  
(b) $\Gamma_2 =     |E_1^0 - E_2^0|$,  
(c) $\Gamma_2 = 1.2 |E_1^0 - E_2^0|$.}
\end{figure}

We consider two states with energies $E_1^0$ and $E_2^0$, which are 
coupled to the continuum channel. This coupling induces an
effective interaction described by the mass operators 
\begin{eqnarray}
    \left(
      \begin{array}{cc}
      V_{11} & V_{12} \\
      V_{21} & V_{22} 
      \end{array}
    \right)
    & = &
    \left(
      \begin{array}{cc}
      g_1^2   \langle 1|G(E)|1 \rangle &  g_1 g_2 \langle 1|G(E)|2 \rangle \\ 
      g_1 g_2 \langle 2|G(E)|1 \rangle &  g_2^2   \langle 2|G(E)|2 \rangle 
      \end{array}
    \right)
\label{EffInt} 
\end{eqnarray}
where $g_i$ is the coupling between the state $i$ and the continuum,  
$|i\rangle$ is the corresponding vertex, and $G(E)$ is the Green function
of the continuum channel.  
The mass operators $V_{ij}$ are energy dependent: in particular,
they have kinematical cuts starting from the continuum threshold.
When the states 1 and 2 are coupled to the {\it open} channel, they  
are not the eigenstates of the full Hamiltonian, but become resonances
located on the second sheet, where $\mbox{\rm Im} V_{ij} \leq 0$, and 
they can be identified with the poles of the Green function.
Their position is determined by the equation
\begin{eqnarray}
   \det
   \left(
      \begin{array}{cc}
      E - E_1^0 - V_{11} &             V_{12} \\
                  V_{21} & E - E_2^0 - V_{22} 
      \end{array}
    \right)
    = 0
\label{EEC}
\end{eqnarray}
For the sake of simplicity, we assume the states 1 and 2 to be far above
the threshold and neglect the energy dependence of the mass operators. 
Furthermore we consider only the imaginary parts of the mass
operators\footnote{The real parts lead to the well known shifts due to
the diagonal interactions and to the mutual repulsion due to the
nondiagonal interactions.} and use the following parametrization:
$V_{11}=-i\Gamma_1/2$, $V_{12}=V_{21}=-i\sqrt{\Gamma_1\Gamma_2}/2$,
$V_{22}=-i\Gamma_2/2$. 
Equation (\ref{EEC}) is then reduced to  
\begin{eqnarray}
   \det
   \left(
      \begin{array}{cc}
      E - E_1^0 - i \Gamma_1/2    & -i\sqrt{\Gamma_1\Gamma_2}/2  \\
      -i\sqrt{\Gamma_1\Gamma_2}/2 & E - E_2^0 - i \Gamma_2/2  
      \end{array}
    \right)
    = 0
\label{EECa}
\end{eqnarray}
which has the resonance pole solutions:  
\begin{eqnarray}
    E_{\pm} = \frac{(E_1^0+E_2^0)-i(\Gamma_1+\Gamma_2)/2}{2} \pm
    \frac{1}{2}
    \sqrt{(E_1^0-E_2^0-i\Gamma_1+i\Gamma_2)^2 - \Gamma_1\Gamma_2} 
\end{eqnarray}
Considering the resonance positions as functions of the individual
couplings ($\Gamma_1\sim g_1^2$, $\Gamma_2\sim g_2^2$) we find the
different types of solutions shown in Fig.\ref{TSP}.
The trajectories plotted correspond to the solutions $E_{\pm}$ for variable
width $\Gamma_1$ and fixed values of $\Gamma_2$ and the
initial positions $E_1^0$ and $E_2^0$. 
As the width $\Gamma_1$ increases starting from zero, the two poles
first attract each other, but then the attraction turns into repulsion.   
There are three different types of solutions for the trajectories
$E_{\pm}(\Gamma_1)$, which correspond to the following cases: 
\begin{eqnarray}
   (a) & & \Gamma_2  <  |E_1^0 - E_2^0|  \nonumber \\ 
   (b) & & \Gamma_2  =  |E_1^0 - E_2^0|  \label{ST} \\ 
   (c) & & \Gamma_2  >  |E_1^0 - E_2^0|  \nonumber 
\end{eqnarray}
In the case (a) the trajectory corresponding to the solution
$E_{-}(\Gamma_1)$ remains bounded, as $\Gamma_1$ goes to infinity,
while in the case (c) the bounded trajectory corresponds to the
solution $E_{+}(\Gamma_1)$. In the case (b) the trajectories
collide in the complex plane at $\Gamma_1=\Gamma_2$. 
In all cases, infinite growth of $\Gamma_1$ does not make both states
infinitely broad: there is always a solution with width 
smaller than $|E_1^0 - E_2^0|$. 

\clearpage


\end{document}